\documentclass[aps,prl,superscriptaddress,reprint]{revtex4-2}   

\usepackage{graphicx}
\usepackage{dcolumn}
\usepackage{bm}
\usepackage{gensymb}

\usepackage{hyperref}
\hypersetup{
    colorlinks=true,
    linkcolor=blue,
    filecolor=magenta,      
    urlcolor=cyan,
    }

\begin{document}
\title{Ferroelastic twin angles at the surface of CaTiO\textsubscript{3} quantified by PhotoEmission Electron Microscopy}

\author{G. Magagnin}
\author{C. Lubin}
\affiliation{SPEC, CEA, CNRS, Universit\'{e} Paris-Saclay, CEA Saclay, 91191 Gif-sur-Yvette, France.}

\author{M. Escher}
\author{N. Weber}
\affiliation{Focus GmbH, Neukirchner Straße 2, H\"{u}nstetten-Kesselbach, D-65510, Germany.}

\author{L. Tortech}
\affiliation{NIMBE, CEA, CNRS, Universit\'{e} Paris-Saclay, CEA Saclay, 91191 Gif-sur-Yvette, France}

\author{N. Barrett}
\email{nick.barrett@cea.fr}
\affiliation{SPEC, CEA, CNRS, Universit\'{e} Paris-Saclay, CEA Saclay, 91191 Gif-sur-Yvette, France.}

\date{\today}

\begin{abstract}
We use photoemission electron microscopy to measure the ferroelastic twin wall angles at the surface of CaTiO$_\mathrm{3}$(001) and deduce the strain ordering. We analyze the angular dependence of the photoelectron emission from different domain surfaces, each with their own characteristic tilt angle in the factory roof-like topography. By considering the surface topography as a field perturbation, the offset in the photoemission threshold can be directly related to the tilt angles. With knowledge of the symmetry allowed twin walls we quantify twin topography between 179.1\degree{} to 180.8\degree{}.
\end{abstract}

\maketitle

At twin boundaries in ferroelastic materials, the spontaneous strain changes sign over only a few nanometers~\cite{Hayward1997} giving rise to strong gradients which can generate new properties quite distinct from those of the adjacent domains. Superconductivity~\cite{Aird1998}, polarity~\cite{VanAert2012, Goncalves-Ferreira2008} and chirality~\cite{Goncalves-Ferreira2008} have all been reported in twin walls. Such emerging functionalities are absent in the bulk~\cite{Salje2009, Viehland2014} and provide a new perspective of “the material is the machine”~\cite{Bhattacharya2005}. In addition, their nanometric dimensions make them potentially 2D functional objects.

The polar character of twin walls was predicted theoretically~\cite{Janovec2006} and simulations suggest that twin wall polarity in CaTiO$_\mathrm{3}$ and SrTiO$_\mathrm{3}$~\cite{Scott2012} can be switched by an applied field~\cite{Zykova-Timan2014}. If this were the case then ferroelastic materials with ferroelectric twin walls would be promising for robust, high-density information storage. Harnessing such functionality requires controlling twin wall polarity at the surface which, in turn, depends on the strain state of the domain twins.

Twinning gives rise to surface topography with a characteristic factory roof-like structure. Each twin has a distinct angle at the surface, often within a degree of 180\degree{} (flat surface) defined by the strain tensor compatibility across the wall~\cite{Yokota2014}. The tensors in turn define the local strain gradients and therefore directly influence both the magnitude of the wall polarity and, potentially, the switching field.  Novak and Salje studied the distribution of lattice strain near the intersection of surface layers and twin boundaries~\cite{Novak1998}. They found that twin boundaries close to the surface show a groove-ridge profile on each side of the twin boundary which are expected to generate local polarization via the flexoelectric~\cite{Zubko2013} (or other) coupling effects~\cite{Salje2016flexo}. Quantification of the twin angles at the surface of ferroelastic materials is therefore an essential step towards demonstrating polarity switching in ferroelastic twin boundaries and understanding the electromechanical coupling between strain and polarity.

CaTiO$_\mathrm{3}$ is the archetypal perovskite, ferroelastic below 1150°C with a Pbnm orthorhombic structure. It consists of corner-linked TiO$_\mathrm{6}$ octahedra with Ca atoms sitting in between, distorted from the ideal cubic perovskite by two independent octahedral tilts written as a$^\mathrm{-}$a$^\mathrm{-}$c$^\mathrm{+}$ in Glazer notation~\cite{Glazer1972}. By symmetry, one of the tilts goes to zero at the twin wall, allowing for the emergence of a competing secondary order parameter~\cite{Goncalves-Ferreira2008}. Biquadratic coupling between the primary and secondary order parameter yields two equivalent ground states for the wall polarity~\cite{Salje2016flexo}. However, the flexoelectric induced strong polarization~\cite{Stengel2013} may break inversion symmetry and favor a specific polarization direction in the twin wall.

Twin walls in CaTiO$_\mathrm{3}$ have been studied using aberration-corrected transmission electron microscopy~\cite{VanAert2012}. Second harmonic generation (SHG) provides another proof of the loss of inversion symmetry~\cite{Yokota2014} but with a spatial resolution limited to 0.5 $\mu$m. Eliseev \textit{et al.} have carried out a theoretical study of the DW/surface intersection in CaTiO$_\mathrm{3}$~\cite{Eliseev2012}. However, little work exists on the direct measurement of the twin angles. Electron imaging of charged (ferroelectric) surfaces was proposed by Le Bihan~\cite{Bihan1989} and successfully applied to visualize ferroelastic domains in barium titanate while low energy electron microscopy has yielded valuable data on CaTiO$_\mathrm{3}$ surface topography and structure~\cite{Nataf2017, Zhao2019}.

\begin{figure*}[ht]
\includegraphics{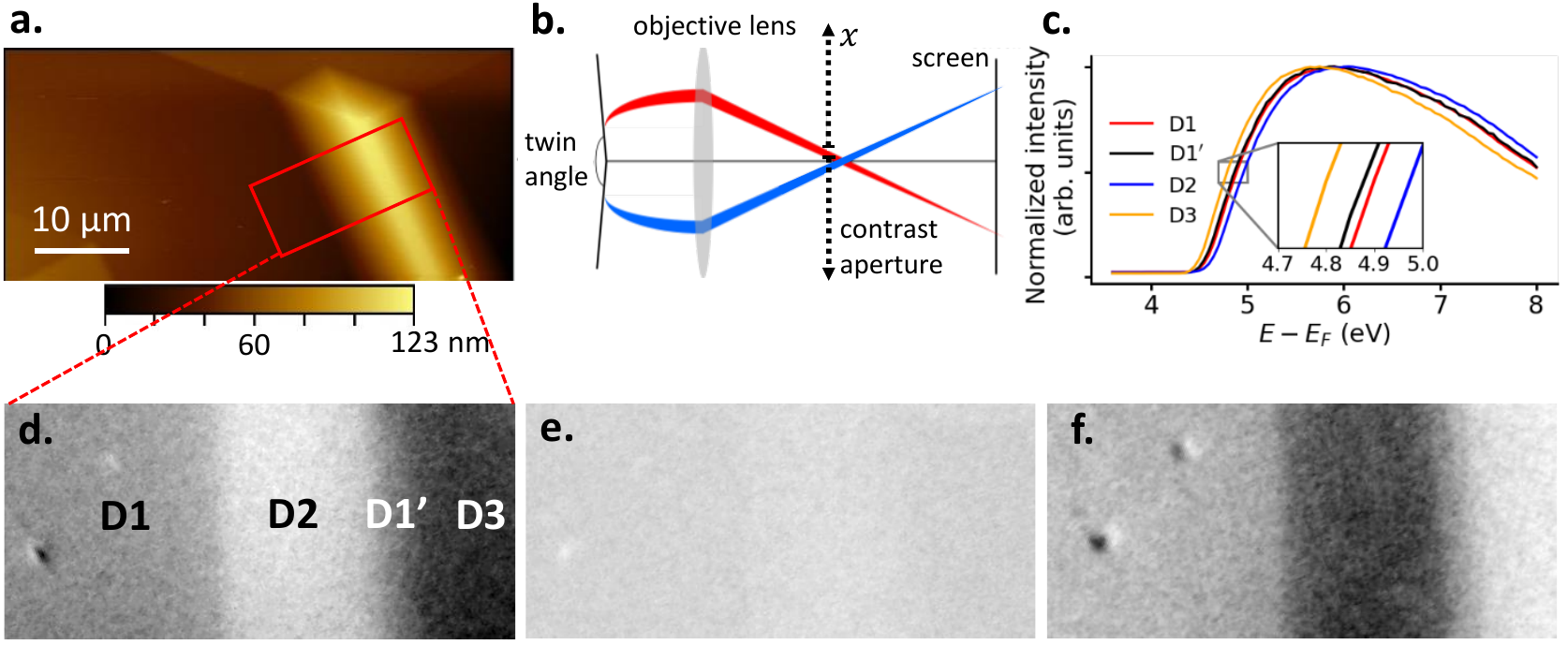}
\caption{a. AFM topography image of the CaTiO$_\mathrm{3}$ surface with a red box highlighting the area of interest containing domains D$_\mathrm{1}$, D$_\mathrm{2}$, D$_\mathrm{1'}$ and D$_\mathrm{3}$. The surface topography is visible with the twin D$_\mathrm{2}$/D$_\mathrm{3}$ on the right hand side. The domains of interest  (b) Schematic showing the angular selection by the contrast aperture in the back focal plane of photoelectron emission from twin domains, here the electron emission in red is favoured (c) Photoemission threshold spectra from domains D$_\mathrm{1}$, D$_\mathrm{2}$ and D$_\mathrm{3}$ (d-f) PEEM images acquired at $E-E_f$ = 4.3 eV for CA positions +140, 0 and -140 $\mu$m with respect to the optical axis.}
\label{fig1}
\end{figure*}

Energy-filtered PhotoEmission Electron Microscopy (PEEM) is a non-destructive, surface-sensitive imaging technique with a high spatial and energy resolution. Contrast in PEEM arises from local chemistry, work function~\cite{Escher2010}, electrical or physical topography~\cite{Nepijko2001, Lavayssiere2013}. We have developed a quantitative approach using the specificities of photoemission electron microscopy to determine the twin angle present at the surface of CaTiO$_\mathrm{3}$ thanks to a simple model of the imaging electron optics. This has been done by exploring the angular space of the PEEM images, specifically, electrons cross the diffraction plane on the optical axis for normal emission and off-axis for off-normal emission. By positioning an aperture in the back focal plane, a given angular range can be selected to quantify the twin angles.

Optical microscopy and atomic force microscopy (AFM) are also sensitive to the surface topography, however, the lateral resolution of PEEM ($\sim$50 nm) is much better than optical microscopy while AFM is a scanning technique. The high resolution of parallel imaging of PEEM also opens the perspective of studying ferroelastic domain dynamics.

The sample is a CaTiO$_\mathrm{3}$ (001) single crystal from SurfaceNet GmbH. Before introduction into the vacuum system, the sample was exposed for 5 min to ozone at room temperature to remove the organic contamination. Annealing at 650°C in vacuum is used to desorb the oxidized contaminants and produces near surface oxygen vacancies, helping to alleviate charging problems during the photoemission process~\cite{Nataf2017}. Experiments were carried out at 300°C to further avoid charging. A focused He I source (21.2 eV) was used in a ScientaOmicron NanoESCA II PEEM. Photoemission threshold image series are acquired as a function of the electron energy with respect to the sample holder Fermi level, $E-E_F$, in steps of 50 meV and with an energy resolution of 100 meV, as determined by the analyzer slit and pass energy. Images are normalized with respect to the signal from a homogeneous area of the sample in order to account for detector inhomogeneities. The image series were analyzed using an automatic procedure fitting the pixel-by-pixel threshold spectra with an error function~\cite{Barrett2013}. The non-isochromaticity in the vertical direction due to the dispersion in the hemispherical imaging analyzer is corrected~\cite{Barrett2013}. The contrast aperture (CA) in the back focal plane has a diameter of 150 $\mu$m. Image series were acquired for CA positions between -230 $\mu$m and + 230 $\mu$m in steps of 10 $\mu$m with respect to the optical axis.

Complementary AFM was performed using a Nano-Observer (CS-Instruments) in contact mode with FORTA tips (AppNano) with stiffness of 1.6 N/m to characterize the surface without scratching.

The effect of the CA is shown in the schematic of Fig.~(\ref{fig1}b). Higher off-centering of the CA improves dramatically the domain topography contrast in PEEM thanks to the angular selection but also induces a shift of the energy scale. Off-normal electrons have velocity components perpendicular and parallel to the sample surface, as a result, the kinetic energy measured inside the PEEM will be slightly lower, and the threshold for photoemission is shifted to higher energy within the reference frame of the PEEM. 

Emission from domains with different tilt angles are centered at different positions in the diffraction plane, giving rise to intensity variation as shown in Fig.~\ref{fig1}d-f via the angular selection by the CA (Fig.\ref{fig1}b). We focus on the domains labeled D$_\mathrm{1}$, D$_\mathrm{2}$, D$_\mathrm{1'}$ and D$_\mathrm{3}$. Domain D$_\mathrm{1}$ is used for PEEM electron optics alignment and the surface normal coincides with the PEEM optical axis. D$_\mathrm{2}$, D$_\mathrm{3}$ and D$_\mathrm{1'}$ have finite tilt angles with respect to D$_\mathrm{1}$. The twin wall is vertical, therefore, by off-centering the CA horizontally we selectively analyze photoelectrons emitted from domains (Fig.\ref{fig1}d and f) on either side of a twin boundary. When the CA is centered on the optical axis, the contrast between the twin domains is almost zero. In this configuration, the angular difference with respect to D$_\mathrm{1}$ is minimized as in Fig.\ref{fig1}e. Figure~\ref{fig1}(c) shows the spectra for each domain extracted from the threshold image series.

\begin{figure}
\includegraphics[width=8.5cm]{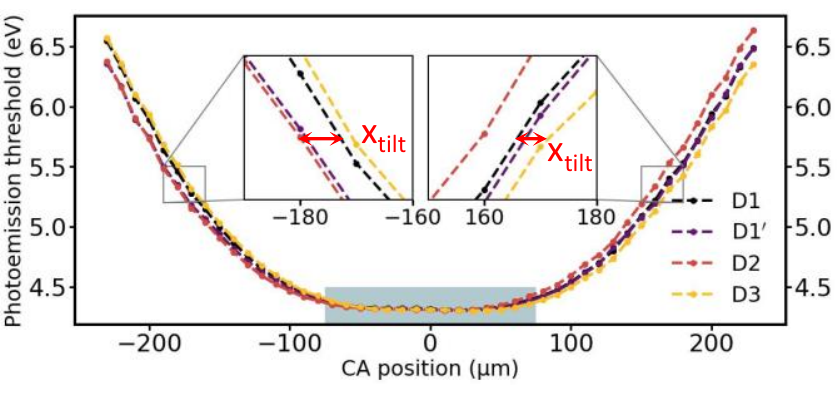}
\caption{Photoemission threshold in domains D$_\mathrm{1}$ to D$_\mathrm{3}$ with the contrast aperture off-centered from -230 $\mu$m to +230 $\mu$m. The insets show the rigid threshold energy shift depending on the surface ferroelastic domains D$_\mathrm{1}$ to D$_\mathrm{3}$. The grey shaded area represents the 150 $\mu$m CA.}
\label{fig2}
\end{figure}

The threshold values are obtained by performing a pixel-by-pixel fit to the spectra with an error function, giving a map of threshold values, as detailed in Supp mat. Figure~\ref{fig2} shows the evolution of the photoemission threshold for D$_\mathrm{1}$, D$_\mathrm{2}$, D$_\mathrm{1'}$, and D$_\mathrm{3}$ of Fig.\ref{fig1} as a function of the CA position from –230 $\mu$m to +230 $\mu$m.

\begin{figure}
\includegraphics[width=8.5cm]{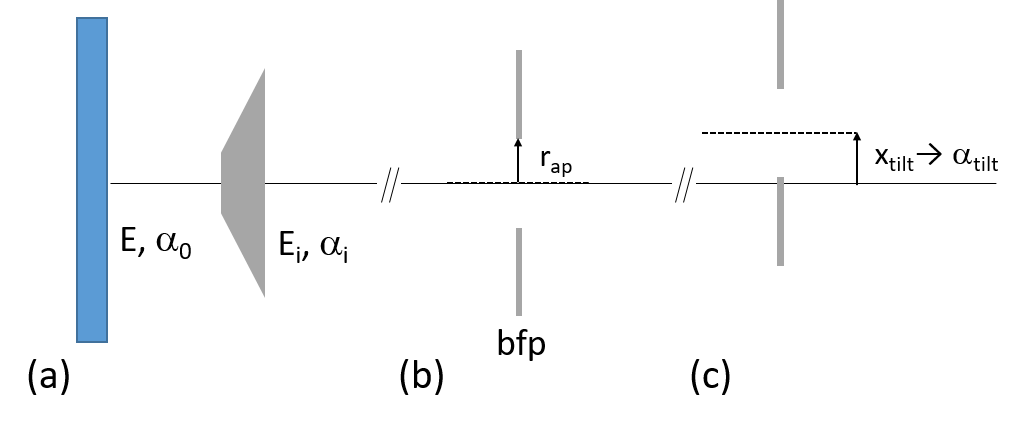}
\caption{(a) electrons are emitted with energy E and at an angle $\alpha$ in the laboratory reference frame are transported at E$_\mathrm{i}$ in the PEEM making an angle $\alpha_\mathrm{i}$ with the optical axis. (b) CA of radius r$_\mathrm{ap}$ centered in the objective lens back focal plane (c) CA off-centered at x$_\mathrm{tilt}$ corresponding to the angular deviation in the PEEM due to $\alpha_\mathrm{tilt}$ of the domain twin.}
\label{fig3}
\end{figure}

Close to the optical axis, the measured threshold is constant at 4.05 eV. When the CA is off-centered further than its physical radius ($\sim$75 $\mu$m), electrons on the optical axis are physically blocked, effectively switching to a dark field imaging mode where higher angle emission from one side of the optical axis is enhanced at the expense of emission from the other side. The contrast between domains is enhanced (Fig.\ref{fig1}d,f) not only because the threshold value increase but also because the difference between domain thresholds increases. The threshold energy curves have the same form but they are not centered at the same CA position. D$_\mathrm{2}$ and D$_\mathrm{3}$ have surface tilts of opposite signs and are rigidly shifted respectively to the left and right with respect to that of D$_\mathrm{1}$ and D$_\mathrm{1'}$. When the domain surface is tilted by an angle $\alpha_{tilt}$ the photoelectron intensity in the back focal plane is also off-centered by a distance $x_{tilt}$, experimentally obtained from the centroid of the two parabolic branches in Fig.~\ref{fig2}. Therefore, by measuring the shift in the threshold curves and with a knowledge of the electron optics, it should be possible to quantify the twin angles, as suggested in Fig.~\ref{fig3}.

Phase conservation in the PEEM is given by  

\begin{equation}
\label{eq:1}
\sqrt{E} r_0 sin(\alpha_0) =  \sqrt{E_i} r_i sin(\alpha_i)\\
\end{equation}

where $\alpha_\mathrm{0}$ the emission angle with respect to the sample normal, $E_{i}$ the electron energy in the PEEM column, $\alpha_{i}$ the electron angle with respect to the optical axis in the PEEM.  The objective lens magnification M is defined as $r_i$/$r_0$. For small angles, $sin(\alpha_i) = \frac{x}{l}$ where l is the distance between the CA and the first image plane.  For simplicity, we assume an isotropic electron emission up to 90°, i.e. $\alpha=90^{\circ}$. For a tilted surface, the optical axis in the back focal plane is shifted by x$_\mathrm{tilt}$. The photoemission threshold $E_{Thr}$ is given by equation \ref{eq:2} (details in supplementary materials). 

\begin{eqnarray}
E_{Thr} = \frac{E_iM}{l^2}(x + x_\mathrm{tilt}\pm r_{ap})^2
\label{eq:2}.
\end{eqnarray}

\begin{figure}
\includegraphics{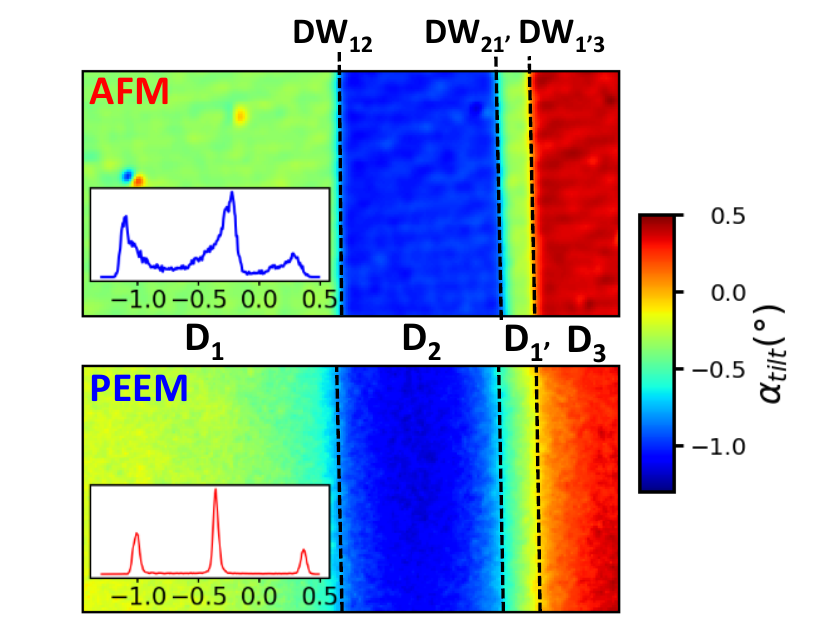}
\caption{PEEM and AFM angular maps of the analyzed area, with an indication of the domain walls. Insets with the histogram of the angles extracted from the AFM and PEEM maps and correspondence with D$_\mathrm{1}$ to D$_\mathrm{3}$.}
\label{fig4}
\end{figure}

The evolution of the photoemission threshold with the CA lateral position is therefore a stretched parabola with a flat central range defined by $r_{ap}$ of constant minimum threshold. In the NanoESCA setup, $l$ = 165 mm, $M$ = 32 and $E_i$ = 2000 eV, which allows to extract $x_{tilt}$ at each pixel. From $x_{tilt}$, we can then work back to the surface tilt angle $\alpha_{tilt}$ by considering a periodic triangular surface topography as a perturbation of the local electric field~\cite{Bok1968} (details in supplementary materials). Equation \ref{eq:3} expresses the relation between the surface tilt angle $\alpha_{tilt}$ and the CA position in the back focal plane $x_{tilt}$.

\begin{eqnarray}
\alpha_{tilt} = c_k \frac{\pi^2}{4}(\frac{\sqrt{2m_eeU_o}}{\hbar} \cdot \sqrt{\frac{L}{d}})^{-1} \cdot x_{tilt}
\label{eq:3}.
\end{eqnarray}

with c$_\mathrm{k}$ is the conversion factor between position in the back focal plane and reciprocal lattice vector for the PEEM settings used here (4.14 $\mathrm{\AA{}}^{-1}\mathrm{mm^{-1}}$), $m_e$ the electron mass, $U_0$ the bias between the sample and extractor (20 kV), $2L = 20 \; \mu{m}$ the surface topography periodicity and $d = 2.5 \; mm$ the distance between the sample and extractor. The angle map is represented in Fig.\ref{fig4}a and compared with that measured by AFM.

There is a good qualitative agreement between the PEEM and AFM maps. The PEEM analysis correctly discriminates the ferroelastic domains D$_\mathrm{1}$, D$_\mathrm{2}$, D$_\mathrm{1'}$ and D$_\mathrm{3}$. Surfaces tilted with a positive or negative angle are revealed and the narrow domain D$_\mathrm{1'}$ between D$_\mathrm{2}$ and D$_\mathrm{3}$, which has the same angle as D$_\mathrm{1}$ is also resolved. The latter has a domain width of $\sim$ 1.7 $\mu$m. It should be noted that the usual method in the PEEM of imaging in the reciprocal space to deduce surface angles would have been impossible for a domain this small since it is beyond the limit of usual field apertures. High resolution real-space imaging to deduce angular maps with the sub-micron resolution is necessary.

The histograms of the  PEEM and AFM angular maps are shown in the insets of Fig.~\ref{fig4}. The overall angular range, as determined by PEEM is between -1.25\degree{} and 0.45\degree{}, in agreement with the known CaTiO$_\mathrm{3}$ twin angles~\cite{Nataf2017}. D$_\mathrm{2}$ and D$_\mathrm{3}$, centered at -1.0\degree{} and +0.4\degree{}, respectively, show good agreement between AFM and PEEM. The main discrepancy between the two angular maps remains the tilt angle of D$_\mathrm{1}$. This is due to the residual alignment offset between PEEM and AFM.

The angular spread for D$_\mathrm{1}$ and D$_\mathrm{2}$ is much smaller in AFM compared to the PEEM, typically 0.05\degree{} compared to 0.15\degree{}. The higher angular spread for the PEEM data is related to the finite aperture size of the electron optics. The finite CA radius allows a spread in acceptance angles and hence in threshold values, contrary to the AFM analysis. A second factor is that the PEEM acquires data by parallel imaging at fixed lens parameters. There is a weak correlation even for microscopic fields of view between position and take-off angle which adds to the angular broadening whereas AFM acquires data sequentially, at each data point measuring the same slope, and is immune to angular cross-talk.

The twin angles calculated from the $\alpha_{tilt}$ values are DW$_\mathrm{12}$ = 179.1\degree{} $\pm$ 0.2\degree{}, DW$_\mathrm{21'}$ = 180.7 $\pm$ 0.2\degree{}, and DW$_\mathrm{1'3}$ = 180.8 $\pm$ 0.2\degree{} and are reported in table \ref{tab:1}.

\begin{table}
\caption{\label{tab:1} Domain wall twin angles and their corresponding spontaneous strain pairs. The uncertainty corresponds to 2$\sigma$.}
\begin{ruledtabular}
\begin{tabular}{cccr}
 & DW$_\mathrm{12}$ & DW$_\mathrm{21'}$ & DW$_\mathrm{1'3}$  \\ \hline
Twin angle & 179.1\degree{} $\pm$ 0.2\degree{} & 180.7 $\pm$ 0.2\degree{} & 180.8 $\pm$ 0.2\degree{} \\
Strain state & S$_\mathrm{vi}$/S$_\mathrm{iii}$ & S$_\mathrm{iii}$/S$_\mathrm{iv}$ & S$_\mathrm{iv}$/S$_\mathrm{v}$ \\
\end{tabular}
\end{ruledtabular}
\end{table}

\begin{figure}
\includegraphics{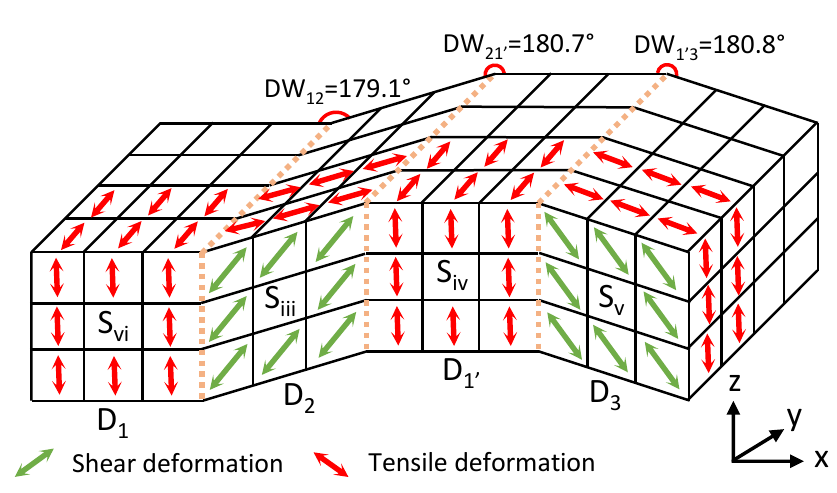}
\caption{Schematic of the spontaneous strain configuration in the analyzed area with its indexed spontaneous domain strains, reported in the supplemental materials.}
\label{fig5}
\end{figure}

There are six possible spontaneous strain orientations in CaTiO$_\mathrm{3}$ which satisfy stain compatibility deduced from symmetry~\cite{Sapriel1975, Yokota2014, Nataf2017}. Given the experimental angles, the twin walls are of type W~\cite{Yokota2014} and are described by x = $\pm$y. From the sequence 179.1\degree{}/180.9\degree{}/180.9\degree{}, we deduce a strain ordering of S$_\mathrm{vi}$/S$_\mathrm{iii}$/S$_\mathrm{iv}$/S$_\mathrm{v}$ (supp mat), as represented in Fig.~\ref{fig5} and summarized in table \ref{tab:1}.

This analysis is limited to domains aligned vertically in the PEEM, i.e. running orthogonal to the lateral displacement of the CA. However, it would be straightforward to extend the methodology to two dimensions to analyze twin structures along all combinations of $<100>$ and $<110>$ by using the full x-y in-plane positions of the CA. As discussed, the finite CA radius gives rise to an angular spread, however, in the limit of small angles, this does not influence the mean twin angles. 

We have used threshold PEEM imaging to measure twin angles at the surface of ferroelastic CaTiO$_\mathrm{3}$  with its characteristic factory roof-like structure. By off-centering the contrast aperture from the optical axis, contrast due to the physical surface topography is enhanced by collecting high angular photoelectrons in a near dark-field mode. Electrons emitted at higher angles have a higher apparent value of the photoemission threshold. Using a model of the electron optics, integrating the perturbation of the extractor field by surface twin topography we can quantify the twin angles, and by comparison with the symmetry allowed twin walls we can deduce directly the surface strain ordering. The results agree well with the independent measurements by AFM. They provide a unique insight into electromechanical coupling responsible for twin wall polarity at the surface and, potentially, a handle to control twin wall polarity. 

\bibliographystyle{apsrev4-1}
\bibliography{references} 

\end{document}


\title{Supplemental materials: Ferroelastic twin angles at the surface of CaTiO\textsubscript{3} quantified by PhotoEmission Electron Microscopy}

\maketitle

\section{Photoelectron kinetic vector}

\begin{figure}[ht]
\centering
\includegraphics[max width=\textwidth/3]{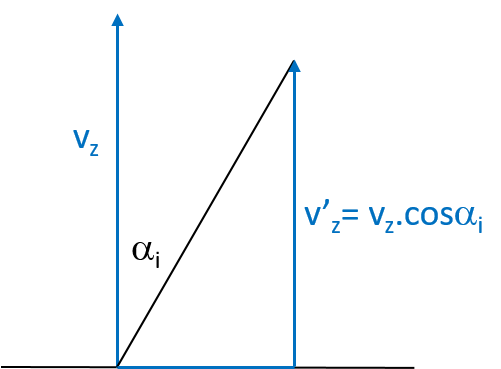}
\caption{Schematic of the electron kinetic vector for a normal emission and an emission with an angle $\alpha_{i}$.}
\label{supp:1}
\end{figure}

Electrons with higher emission angles have lower kinetic energy as measured on the scale of $E-E_F$ making the threshold appear at higher energy, as shown in figure \ref{supp:1}. Electron emission normal to the sample surface is favored when the CA is centered and electrons emitted off-normal are favored when the CA is off-centered. The extractor field is measured perpendicular to the average sample surface so photoelectrons with higher emission angles have lower kinetic energy ($E_k \approx v_k^2$) as measured on the scale of $E-E_F$ and the threshold will appear at higher energy.

\section{Model of the photoemission threshold evolution with CA lateral displacement}

Let an electron emitted at an angle $\alpha_\mathrm{0}$ make an angle $\alpha_\mathrm{i}$ with respect to the PEEM electron optical axis and cross the back focal plane at x from the axis. To extract an $x_{tilt}$ value at each pixel of the analyzed area, we model the electron path in the PEEM. The relevant parameters are the electron kinetic energy $E$, the PEEM column energy $E_{i}$, the magnification in the first image plane $M$, the CA position in the back focal plane $x$.

The phase conservation equation of the photoelectrons is written in equation \ref{suppeq:1}. For simplicity, we assume an isotropic emission of electrons from 0° to 90°, i.e. $\alpha_{0}=90^{\circ}$ and the magnification of the image in the first image plane is given by the ratio between the object size $r_0$ and the first intermediate image size $r_i$. For a tilted surface, the optical axis in the back focal plane is shifted by a distance $x_{tilt}$. 

\begin{equation}
\begin{split}
\label{suppeq:1}
\sqrt{E} r_0 sin(\alpha_0) &=  \sqrt{E_i} r_i sin(\alpha_i)\\
sin(\alpha_i)& = \frac{x}{l}
\end{split}
\end{equation}

The shift in measured photoemission threshold $E_{Thr}$ with the CA position for a surface with tilt angle $\alpha_\mathrm{tilt}$ is then given by equation \ref{suppeq:2}.

\begin{equation}
\begin{split}
\label{suppeq:2}
&\Delta{E_{Thr}} = \frac{E_iM^2}{l^2}(x + r_{ap} - x_{tilt})^2, \:for \:x \leq -r_{ap} + x_{tilt} \\
&\Delta{E_{Thr}} = \frac{E_iM^2}{l^2}(x - r_{ap} - x_{tilt})^2, \:for \:x \geq r_{ap} + x_{tilt}  \\
&E_{Thr} = 0, \:for -r_{ap} + x_{tilt} \leq x \leq r_{ap} + x_{tilt}
\end{split}
\end{equation}

\begin{figure}[ht]
\centering
\includegraphics[max width=\textwidth/2]{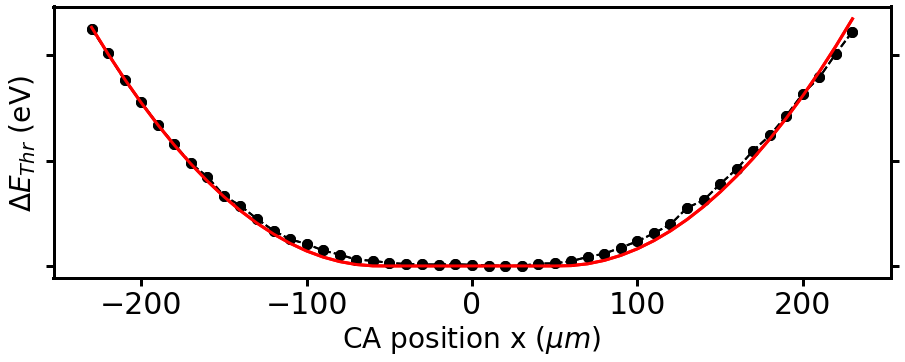}
\caption{Experimental variation in photoemission threshold as a function of the 150 $\mu$m CA position in the diffraction plane from -230 $\mu$m to 230 $\mu$m. In red plot of eqn.\ref{suppeq:1} using $r_{ap}$ = 55 $\mu$m.}
\label{supp:2}
\end{figure}

It is a parabola with a central flat range fixed by $r_{ap}$. In our PEEM set-up $l=165mm$, $M=32$, and $E_i=2000eV$. Fig. \ref{supp:2} shows the experimental data for $x_{tilt}=0$ from a pixel in the domain $D_1$ together with, in red, eqn \ref{suppeq:2}. Note that the best fit is obtained for $r_{ap}$ = 55 $\mu$m rather than the nominal radius of 75 $\mu$m. We fit the increasing threshold energy by two parabolic branches out with $\pm r_{ap}$ and calculate the off-centering from the optical axis $x_{tilt}$ from the midpoint between the two branches.

\section{Model of the surface tilt perturbation at the back focal plane}

\subsection*{Triangle Perturbation}

\begin{figure}[ht]
\centering
\includegraphics[max width=9.5cm]{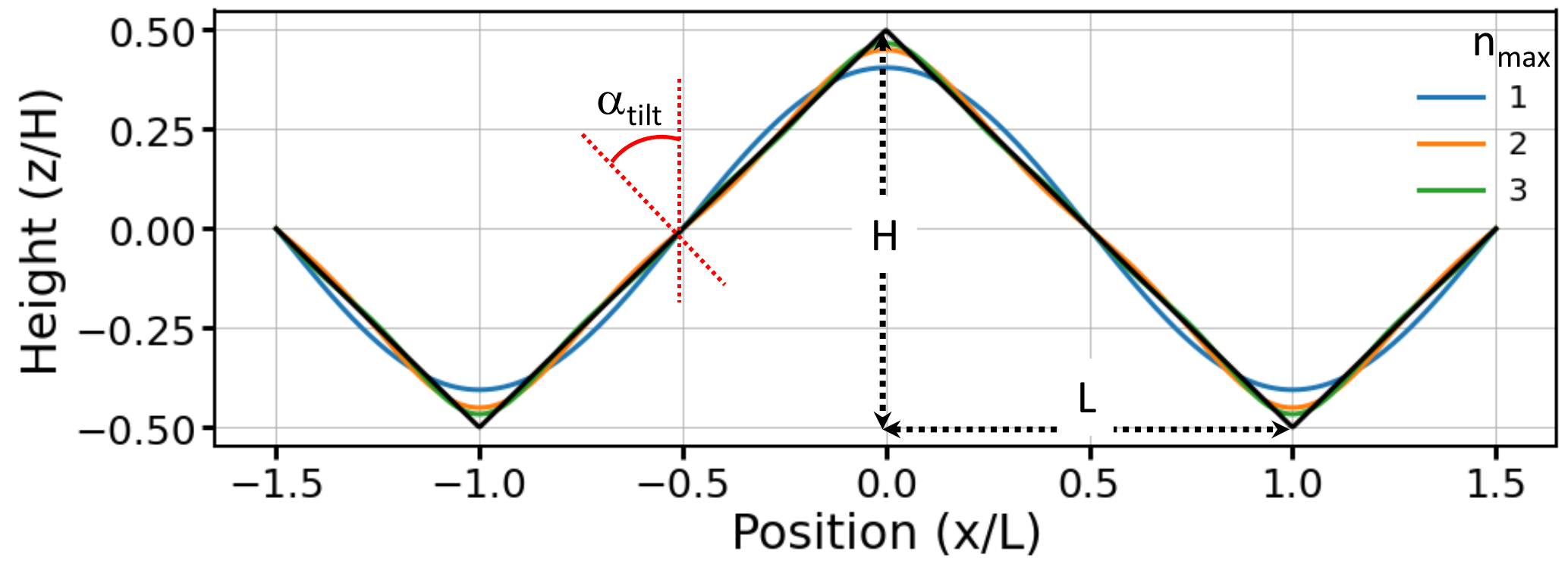}
\caption{Periodic triangular structure simulating ferroelastic topography and representation of the triangular function Fourier series for $n_{max}$ = 1 to 3. The surface is tilted by the angle $\alpha_{tilt}$ to the z-axis.}
\label{supp:3}
\end{figure}

To recover $\alpha_\mathrm{tilt}$ from $x_\mathrm{tilt}$ we model the roof topography of the $CaTiO_3$ surface as a perturbation of the electric field above the surface and solve the equations of motion for the photoemitted electrons, We assume a periodic structure (Triangular Function) with a homogeneous work function of period $2 L$ and amplitude $0.5 \mathrm{H}$. The triangular structure is defined as

$$
z_{p}(x)=\left\{\begin{array}{cc}
H / L\left(x-2 L n+\frac{1}{2} L\right) & x \in[2 L n-L, 2 L n], n \in \mathbb{Z} \\
-H / L\left(x-2 L n-\frac{1}{2} L\right) & x \in[2 L n, 2 L n+L], n \in \mathbb{Z}
\end{array}\right.
$$

With $x$ and $y$ being the coordinates in the surface, $z$ normal to the surface, and $z=0$ being in the surface. The $x$-coordinate is the direction of the periodic structure (see Figure \ref{supp:3}). The tilt angle to the surface normal is

$$
\alpha_{tilt}=\left\{\begin{array}{cc}
\tan ^{-1}\left(\frac{\mathrm{H}}{\mathrm{L}}\right) \approx \frac{\mathrm{H}}{\mathrm{L}} & x \in[2 \operatorname{Ln}-L, 2 L n], n \in \mathbb{Z} \\
\tan ^{-1}\left(-\frac{\mathrm{H}}{\mathrm{L}}\right) \approx-\frac{\mathrm{H}}{\mathrm{L}} & x \in[2 L n, 2 L n+L], n \in \mathbb{Z}
\end{array}\right.
$$
assuming the H<<L.

\subsection*{Fourier Series Triangular Function}
The Triangular Function can be expressed as a Fourier Series:

$$
z_{p}(x)=\frac{4 H}{\pi^{2}} \sum_{n=1}^{\infty} \frac{1}{(2 n-1)^{2}} \cos ((2 n-1) k x)
$$

With $k=2 \pi /(2 L)$. We will fix $n_{max}=1$ for the triangular Fourier series as in the vicinity of ($z=0$,$\frac{x}{L}=\pm0.5$), this is a good approximation of the topography. Higher order terms would be necessary near to the apices which correspond to the twin boundaries.

\subsection*{Potential between Sample and Extractor}

For a flat (unperturbed) sample, the potential can be written as

$$
\Phi_{0}(x, y, z)=-U_{0} \frac{z}{d}
$$

where $U_{0}$ is the voltage and $d$ the distance between sample and objective lens. In the following discussion, we neglect the $y$-coordinate, considering only a two-dimensional problem in x-z. With the factory-roof structure, we get a perturbation potential at $z=0$ which can be written as \cite{Bok1968} :

$$
\Phi_{p}(x, z=0)=\Phi_{p 0}+\sum_{n=1}^{\infty} \Phi_{p n} \cos n k x
$$

This leads to the solution :

$$
\Phi(x, z)=-U_{0} \frac{z}{d}+\Phi_{p 0}+\sum_{n=1}^{\infty}\left[\Phi_{p n} \cos n k x \exp (-n k z)\right]
$$

\subsection*{Boundary conditions}
In order to determine the coefficients $\Phi_{p n}$ we substitute for $z_{p}$ :

$$
\begin{aligned}
& x \in[-L, 0] \\
& \Phi\left(x, z_{p}(x)\right)=0=-\frac{U_{0}}{d} \frac{\mathrm{H}}{\mathrm{L}}\left(x+\frac{1}{2} L\right)+\Phi_{p 0}+\sum_{n=1}^{\infty}\left[\Phi_{p n} \cos n k x \exp \left(-n k \alpha_{tilt}\left(x+\frac{1}{2} L\right)\right)\right] ; \\
& x \in[0, L] \\
& \Phi\left(x, z_{p}(x)\right)=0=\frac{U_{0}}{d} \frac{\mathrm{H}}{\mathrm{L}}\left(x-\frac{1}{2} L\right)+\Phi_{p 0}+\sum_{n=1}^{\infty}\left[\Phi_{p n} \cos n k x \exp \left(n k \alpha_{tilt}\left(x-\frac{1}{2} L\right)\right)\right] ; 
\end{aligned}
$$

In the vicinity of $x \approx-1 / 2 L$ we can neglect the influence of the term

$$
\exp \left(-n k \alpha_{tilt}\left(x+\frac{1}{2} L\right)\right) \approx 1
$$

And expressing the linear term of the potential as Fourier Series

$$
-\frac{U_{0}}{d} \alpha_{tilt}\left(x+\frac{1}{2} L\right)=-\frac{U_{0}}{d} \frac{4 H}{\pi^{2}} \sum_{n=1}^{\infty} \frac{1}{(2 n-1)^{2}} \cos ((2 n-1) k x)
$$

The comparison of coefficients yields

$$
\Phi_{p n}=\frac{U_{0}}{d} \frac{4 H}{\pi^{2}} \frac{1}{(2 n-1)^{2}}
$$

\subsection*{Electric Field}
From the electric potential, we yield the electric field:

$$
\begin{gathered}
\frac{\partial}{\partial \mathrm{z}} \Phi(x, z)=\frac{\partial}{\partial \mathrm{z}} \Phi_{o}(z)+\frac{\partial}{\partial \mathrm{z}} \Phi_{p}(x, z) \\
\frac{\partial}{\partial \mathrm{x}} \Phi(x, z)=\frac{\partial}{\partial \mathrm{x}} \Phi_{p}(x, z)
\end{gathered}
$$

In the $z$ direction we can neglect the perturbation contribution:

$$
E_{z}(z)=\frac{\partial}{\partial \mathrm{z}} \Phi(x, z) \approx \frac{\partial}{\partial \mathrm{z}} \Phi_{o}(z)=-\frac{U_{0}}{d}
$$

For the $x$ direction we have:

$$
\begin{gathered}
E_{x}(x, z)=\frac{\partial}{\partial \mathrm{x}} \Phi_{p}(x, z) \\
=\frac{\partial}{\partial \mathrm{x}} \sum_{n=1}^{\infty}\left[\Phi_{p n} \cos n k x \exp (-n k z)\right] \\
=\sum_{n=1}^{\infty}\left[-n k \Phi_{p n} \sin n k x \exp (-n k z)\right] \\
=\sum_{n=1}^{\infty}\left[-\frac{U_{0}}{d} \frac{4 H}{\pi^{2}} \frac{n k}{(2 n-1)^{2}} \sin n k x \exp (-n k z)\right] \\
=-\frac{U_{0}}{d} \frac{4 H}{\pi^{2}} k \sum_{n=1}^{\infty}\left[\frac{n}{(2 n-1)^{2}} \sin n k x \exp (-n k z)\right]
\end{gathered}
$$

\subsection*{Equations of motion}

We use the classical approach since in the vicinity of the perturbated surface the electrons do not have high kinetic energy:

$$
\begin{gathered}
-e E_{x}(x, z)=m_{e} \frac{\partial^{2} x(z)}{\partial \mathrm{t}^{2}} \\
-e E_{z}(z)=-e\left(-\frac{U_{0}}{d}\right)=m_{e} \frac{\partial^{2} z}{\partial \mathrm{t}^{2}}
\end{gathered}
$$

\subsection*{Initial Conditions}
The momentum of an electron leaving the sample at $t=0$ at $z=0$ and $x=-1 / 2 L$ with $E$ being the kinetic energy when leaving the sample surface, can be written as

$$
\begin{aligned}
& p_{z 0}=\sqrt{2 m_{e} E} \cos \left(\alpha_{0}+\alpha_{tilt}\right) \\
& p_{x 0}=\sqrt{2 m_{e} E} \sin \left(\alpha_{0}+\alpha_{tilt}\right)
\end{aligned}
$$

With the electron mass $m_{e}$, tilt angle $\alpha_{tilt}$ and take off angle $\alpha_{0} \in[-\pi / 2, \pi / 2]$.

\subsection*{Solution for the z-axis}

In $z$ direction integration of the momentum over time yields

$$
\begin{aligned}
p_{z}(t)=m_{e} \frac{\partial z}{\partial \mathrm{t}} & =\sqrt{2 m_{e} E} \cos \left(\alpha_{0}+\alpha_{tilt}\right)-\int_{0}^{t} e E_{z}(z) d t^{\prime} \\
p_{z}(t) & =\sqrt{2 m_{e} E} \cos \left(\alpha_{0}+\alpha_{tilt}\right)+e \frac{U_{0}}{d} t
\end{aligned}
$$

Second integration:

$$
z(t)=\int_{0}^{t} \frac{p_{z}(t)}{m_{e}} d t^{\prime}=\sqrt{2 m_{e} E} \cos \left(\alpha_{0}+\alpha_{tilt}\right) t+2 \frac{e U_{0}}{m_{e} d} t^{2}
$$

For threshold emission we neglect the contribution of the initial electron momentum and work with the approximation:

$$
z(t)=2 \frac{e U_{0}}{m_{e} d} t^{2}
$$

\subsection*{Solution for the $x$-axis}
Therefore the equation of motion in $\mathrm{x}$ direction becomes

$$
\begin{aligned}
m_{e} \frac{\partial^{2} x(z(t))}{\partial \mathrm{t}^{2}} & =-e E_{x}(x, z)=-e\left(-\frac{U_{0}}{d} \frac{4 H}{\pi^{2}} k \sum_{n=1}^{\infty}\left[\frac{n}{(2 n-1)^{2}} \sin n k x(t) \exp (-n k z(t))\right]\right) \\
& =\frac{e U_{0}}{d} \frac{4 H}{\pi^{2}} k \sum_{n=1}^{\infty}\left[\frac{n}{(2 n-1)^{2}} \sin n k x(t) \exp \left(-n k 2 \frac{e U_{0}}{m_{e} d} t^{2}\right)\right]
\end{aligned}
$$

Integration over $\mathrm{t}$ yields the momentum in $\mathrm{x}$ direction:

$$
\begin{aligned}
& p_{x}(t)=m_{e} \frac{\partial x(t)}{\partial \mathrm{t}} \\
& =\sqrt{2 m_{e} E} \sin \left(\alpha_{0}+\alpha_{tilt}\right)+\frac{e U_{0}}{d} \frac{4 H}{\pi^{2}} k \sum_{n=1}^{\infty}\left[\frac{n}{(2 n-1)^{2}} \int_{0}^{t} \sin \left(n k x\left(t^{\prime}\right)\right) \exp \left(-n k 2 \frac{e U_{0}}{m_{e} d} t^{\prime 2}\right) d t^{\prime}\right]
\end{aligned}
$$

\subsection*{Approximation for the $x$-axis}
We restrict the series of the electric field to $n=1$ :

$$
p_{x}(t) \approx \sqrt{2 m_{e} E} \sin \left(\alpha_{0}+\alpha_{tilt}\right)+\frac{e U_{0}}{d} \frac{4 H}{\pi^{2}} k \int_{0}^{t} \sin \left(k x\left(t^{\prime}\right)\right) \exp \left(-k 2 \frac{e U_{0}}{m_{e} d} t^{\prime 2}\right) d t^{\prime}
$$

In the proximity of $x(t)=-1 / 2 L(x(t)=1 / 2 L)$ we can approximate

$$
\begin{array}{ll}
\sin (k x(t)) \approx-1 ; & x(t) \approx-1 / 2 L \\
\sin (k x(t)) \approx 1 ; & x(t) \approx 1 / 2 L
\end{array}
$$

Then, the momentum in $x$-direction simplifies to

$$
p_{x}(t) \approx \sqrt{2 m_{e} E} \sin \left(\alpha_{0}+\alpha_{tilt}\right)+\alpha_{tilt} \frac{e U_{0}}{d} \frac{4}{\pi} \int_{0}^{t} \exp \left(-k 2 \frac{e U_{0}}{m_{e} d} t^{\prime 2}\right) d t^{\prime}
$$

Integration of the Gaussian can be approximated to $t \rightarrow \infty$ :

$$
\int_{0}^{\infty} \exp \left(-k 2 \frac{e U_{0}}{m_{e} d} t^{2}\right) d t=\frac{2}{\sqrt{\pi}} \sqrt{\frac{m_{e} d}{2 e U_{0} k}}=\frac{2}{\pi} \sqrt{\frac{m_{e} d L}{2 e U_{0}}}
$$

Hence, the momentum in $x$-direction can be written as

$$
\begin{aligned}
& p_{x} \approx \sqrt{2 m_{e} E} \sin \left(\alpha_{0}+\alpha_{tilt}\right)+\alpha_{tilt} \frac{e U_{0}}{d} \frac{4}{\pi} \frac{2}{\pi} \sqrt{\frac{m_{e} d L}{2 e U_{0}}} \\
& p_{x} \approx \sqrt{2 m_{e} E} \sin \left(\alpha_{0}+\alpha_{tilt}\right)+\alpha_{tilt} \frac{4}{\pi^{2}} \sqrt{2 m_{e} e U_{0}} \sqrt{\frac{L}{d}}
\end{aligned}
$$

The wave vector $k_{x}=p_{x} / \hbar$ in $x$-direction follows as

$$
k_{x} \approx \frac{\sqrt{2 m_{e} E}}{\hbar} \sin \left(\alpha_{0}+\alpha_{tilt}\right)+\alpha_{tilt} \frac{4}{\pi^{2}} \frac{\sqrt{2 m_{e} e U_{0}}}{\hbar} \sqrt{\frac{L}{d}}
$$

The wave vector depends on two terms where the first term has a weak dependence on the surface tilt ($\alpha_{tilt}$ is in the order of 1\degree{}) and the second term depends on the surface tilt and period $2L$ of the triangular topography. We can approximate the topography perturbation defined by the surface tilt $\alpha_{tilt}$ on the wave vector by approximating the first term as constant. The wave-vector difference from the surface topography, $\Delta{k_{x}}$, is expressed as:

$$
\Delta{k_{x}} \approx \alpha_{tilt} \frac{4}{\pi^{2}} \frac{\sqrt{2 m_{e} e U_{0}}}{\hbar} \sqrt{\frac{L}{d}}
$$

We use the factor $c_k=4.14 \: \mathring{A}^{-1}.mm^{-1}$, which converts in the PEEM an angular distance in the reciprocal space into a shift in the back focal plane, defined as $\Delta{k_{x}} = c_k \cdot x_{tilt}$. We can then express the surface domain tilt angle $\alpha_{tilt}$ from the CA shift in the back focal plane $x_{tilt}$:

$$
\alpha_{tilt} = c_k\frac{\pi^2}{4}(\frac{\sqrt{2m_eeU_o}}{\hbar} \cdot \sqrt{\frac{L}{d}})^{-1} \cdot x_{tilt}
$$

\section{Standard deviation}

\begin{figure}[ht]
\centering
\includegraphics[max width=\textwidth/2]{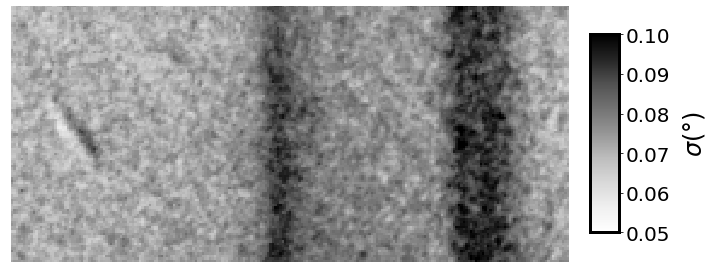}
\caption{Map of the standard deviation $\sigma$ of $\alpha_{tilt}$ as determined by PEEM.}
\label{supp:6}
\end{figure}

The map of the standard deviation $\sigma$ of $\alpha_\mathrm{tilt}$ is shown in figure \ref{supp:6}. From the $\sigma$ map, the calculated $\alpha_{tilt}$ error is approximately of $\pm$ 0.07° in the domains with an error extrema of $\pm$ 0.1° around the ridge/valley domain walls. This can be understood since twin walls mark the transition between two tilt angles, locally modifying the field lines giving rise to a larger spread of angular values.

\section{Possible strain configurations in CaTiO$_\mathrm{3}$}

There are six possible strain states $S_{i}$ in orthorhombic $CaTiO_3$ which are derived from the strain tensor compatibility across the symmetry allowed twin walls \cite{Sapriel1975, Yokota2014}.
\begin{align}
\begin{split}
\label{eq:III2}
S_{i}: \begin{pmatrix} +0.00057 & +0.0055 & 0 \\ +0.0055 & +0.00057 & 0 \\ 0 & 0 & -0.00114 \end{pmatrix}  \\
S_{ii}: \begin{pmatrix} +0.00057 & -0.0055 & 0 \\ -0.0055 & +0.00057 & 0 \\ 0 & 0 & -0.00114 \end{pmatrix}  \\
S_{iii}: \begin{pmatrix} +0.00057 & 0 & -0.0055 \\ 0 & -0.00114 & 0 \\ -0.0055 & 0 & +0.00057 \end{pmatrix}  \\
S_{iv}: \begin{pmatrix} -0.00114 & 0 & 0 \\ 0 & +0.00057 & +0.0055 \\ 0 & +0.0055 & +0.00057 \end{pmatrix}  \\
S_{v}: \begin{pmatrix} +0.00057 & 0 & +0.0055 \\ 0 & -0.00114 & 0 \\ +0.0055 & 0 & +0.00057 \end{pmatrix}  \\
S_{vi}: \begin{pmatrix} -0.00114 & 0 & 0 \\ 0 & +0.00057 & -0.0055 \\ 0 & -0.0055 & +0.00057 \end{pmatrix} 
\end{split}
\end{align}

From the possible spontaneous strain directions, we can deduce the possible domain pairs ($S_{a},S_b$) and calculate the difference in their spontaneous strain tensors $\Delta\epsilon = \epsilon_a-\epsilon_b$. For each pair, the equations of the two possible compatible domain walls, one for each compatible orientation, are calculated from the mechanical compatibility equation $\Delta\epsilon_{ab} x_a x_b = 0$ \cite{Sapriel1975, Yokota2014}. The possible strain pairs are regrouped in the following table \ref{tab:1}.
\newline

\definecolor{Silver}{rgb}{0.752,0.752,0.752}
\begin{longtblr}[
  caption = {Domain wall plane equations for all possible combinations of strain tensors, wall angle with respect to (001), and twin angle.},
  label = {tab:1},
]{
  width = \linewidth,
  colspec = {Q[171]Q[404]Q[108]Q[110]Q[142]},
  row{1} = {Silver},
  row{2} = {Silver},
  row{3} = {c},
  row{4} = {c},
  row{5} = {c},
  row{6} = {c},
  row{7} = {c},
  row{8} = {c},
  row{9} = {c},
  row{10} = {c},
  row{11} = {c},
  row{12} = {c},
  row{13} = {c},
  row{14} = {c},
  row{15} = {c},
  row{16} = {c},
  row{17} = {c},
  row{18} = {c},
  row{19} = {c},
  row{20} = {c},
  row{21} = {c},
  row{22} = {c},
  row{23} = {c},
  row{24} = {c},
  row{25} = {c},
  row{26} = {c},
  row{27} = {c},
  row{28} = {c},
  row{29} = {c},
  row{30} = {c},
  row{31} = {c},
  row{32} = {c},
  cell{1}{1} = {c},
  cell{1}{2} = {c},
  cell{1}{4} = {c},
  cell{1}{5} = {c},
  cell{2}{1} = {c},
  cell{2}{2} = {c},
  cell{2}{4} = {c},
  cell{2}{5} = {c},
  cell{3}{1} = {r=2}{},
  cell{5}{1} = {r=2}{},
  cell{7}{1} = {r=2}{},
  cell{9}{1} = {r=2}{},
  cell{11}{1} = {r=2}{},
  cell{13}{1} = {r=2}{},
  cell{15}{1} = {r=2}{},
  cell{17}{1} = {r=2}{},
  cell{19}{1} = {r=2}{},
  cell{21}{1} = {r=2}{},
  cell{23}{1} = {r=2}{},
  cell{25}{1} = {r=2}{},
  cell{27}{1} = {r=2}{},
  cell{29}{1} = {r=2}{},
  cell{31}{1} = {r=2}{},
  vlines,
  hline{1,3,5,7,9,11,13,15,17,19,21,23,25,27,29,31,33} = {-}{},
  hline{4,6,8,10,12,14,16,18,20,22,24,26,28,30,32} = {2-5}{},
}
             &                                      & Azimuthal   & Inclinaison & Twin            \\
Pairs        & Wall equation                        & angle/[100] & angle/(001) & angle           \\
$S_i/S_{ii}$   & $x=0$                                & 90°         & 90°         & 180°            \\
             & $y=0$                                & 180°        & 90°         & 180°            \\
$S_i/S_{iii}$  & $y=-z$                               & 180°        & 45°         & 180°            \\
             & $3\epsilon_11 (y-z)+2\epsilon_12x=0$ & 98.8°       & 98.7°       & 179.4° (180.6°) \\
$S_i/S_{iv}$   & $z=x$                                & 90°         & 135°        & 180°            \\
             & $3\epsilon_11 (z+x)+2\epsilon_12y=0$ & 171.2°      & 81.3°       & 180.6° (179.4°) \\
$S_i/S_{v}$    & $y=z$                                & 180°        & 135°        & 180°            \\
             & $3\epsilon_11 (y+z)+2\epsilon_12x=0$ & 98.8°       & 81.3°       & 180.6° (179.4°) \\
$S_i/S_{vi}$   & $z=-x$                               & 90°         & 45°         & 180°            \\
             & $3\epsilon_11 (z-x)-2\epsilon_12y=0$ & 8.8°        & 81.3°       & 180.6° (179.4°) \\
$S_{ii}/S_{iii}$ & $y=z$                                & 180°        & 135°        & 180°            \\
             & $3\epsilon_11 (y+z)-2\epsilon_12x=0$ & 98.8°       & 81.3°       & 180.6° (179.4°) \\
$S_{ii}/S_{iv}$  & $z=-x$                               & 90°         & 45°         & 180°            \\
             & $3\epsilon_11 (z-x)+2\epsilon_12y=0$ & 171.2°      & 81.3°       & 180.6° (179.4°) \\
$S_{ii}/S_{v}$   & $y=-z$                               & 180°        & 45°         & 180°            \\
             & $3\epsilon_11 (y-z)-2\epsilon_12x=0$ & 98.8°       & 98.7°       & 179.4° (180.6°) \\
$S_{ii}/S_{vi}$  & $z=x$                                & 90°         & 135°        & 180°            \\
             & $3\epsilon_11 (z+x)-2\epsilon_12y=0$ & 8.8°        & 81.3°       & 180.6° (179.4°) \\
$S_{iii}/S_{iv}$ & $x=-y$                               & 135°        & 90°         & 180.9° (179.1°) \\
             & $3\epsilon_11 (x-y)-2\epsilon_12z=0$ & 45°         & 167.6°      & 180°            \\
$S_{iii}/S_{v}$  & $x=0$                                & 90°         & 90°         & 181.3° (178.7°) \\
             & $z=0$                                & -           & 0°          & -               \\
$S_{iii}/S_{vi}$ & $x=y$                                & 45°         & 90°         & 180.9° (179.1°) \\
             & $3\epsilon_11 (x+y)-2\epsilon_12z=0$ & 135°        & 167.6°      & 180°            \\
$S_{iv}/S_{v}$   & $x=y$                                & 45°         & 90°         & 180.9° (179.1°) \\
             & $3\epsilon_11 (x+y)+2\epsilon_12z=0$ & 135°        & 12.4°       & 180°            \\
$S_{iv}/S_{vi}$  & $y=0$                                & 180°        & 90°         & 178.7° (181.3°) \\
             & $z=0$                                & -           & 0°          & -               \\
$S_{v}/S_{vi}$   & $x=-y$                               & 135°        & 90°         & 179.1° (180.9°) \\
             & $3\epsilon_11 (x-y)+2\epsilon_12z=0$ & 45°         & 12.4°       & 180°            
\end{longtblr}

Note that for each pairing, two angles are possible, the supplementary angle is given in brackets in the table. 

\section{Tilt angle map and twin angles at another area on the sample with a different ferroelastic ordering}

\begin{figure}[ht]
\centering
\includegraphics[max width=\textwidth/2]{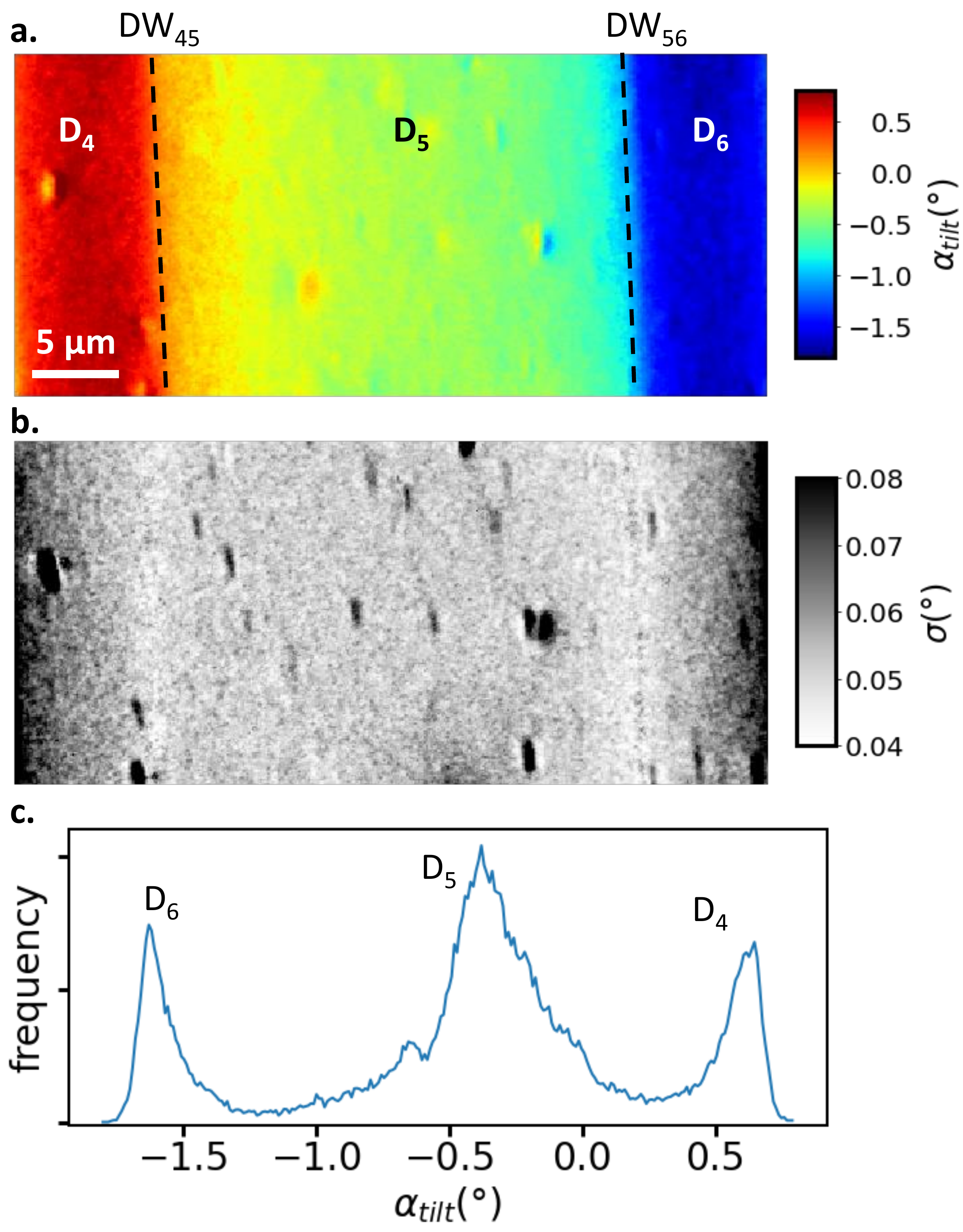}
\caption{(a) $\alpha_\mathrm{tilt}$ map determined by PEEM in another area of the same sample, showing surface twin topography. (b) Map of $\sigma$ of $\alpha_{tilt}$ and (c) Histogram of the $\alpha_{tilt}$ values.}
\label{supp:7}
\end{figure}

 We show in figure \ref{supp:7} a map of the domain tilts in another area of the same sample. Three domains are identified, $D_4$ to $D_6$ with $D_4$ and $D_6$ having tilt angles of opposite signs. The surface tilt values have a standard deviation $\sigma$, reported in figure \ref{supp:7}, of $\sim$ 0.05°. $D_4$ has the same tilt as $D_3$ in the area in the main manuscript. The angles calculated from the angular map are reported in table \ref{tab:2}. Within experimental error, the twin angles are identical to those in the main manuscript. We propose two spontaneous strain configurations corresponding to the measured angles: $S_{v}/S_{iv}/S_{iii}$ or $S_{v}/S_{vi}/S_{iii}$.

\definecolor{Silver}{rgb}{0.752,0.752,0.752}
\begin{table}[ht]
\centering
\caption{Twin angles calculated from the PEEM measurements in the second area. The uncertainty comes from 2$\sigma$.}
\label{tab:2}
\begin{tblr}{
  width = \linewidth,
  colspec = {Q[246]Q[331]Q[319]},
  cells = {c},
  cell{1}{1} = {c=2}{0.577\linewidth},
  cell{1}{3} = {Silver},
  cell{2}{1} = {r=2}{Silver},
  cell{4}{1} = {r=4}{Silver},
  vlines,
  hline{1-2,4,8} = {-}{},
  hline{3,5-7} = {2-3}{},
}
                                        &           & Twin angle (°) \\
{PEEM \\measured}                       & $DW_{45}$    & 179.1 $\pm$ 0.16          \\
                                        & $DW_{56}$    & 179.2 $\pm$ 0.16        \\
{Possible\\theoretical\\W type \\walls} & $S_{iii}/S_{iv}$ & 179.1          \\
                                        & $S_{iii}/S_{vi}$ & 179.1          \\
                                        & $S_{iv}/S_{v}$ & 179.1          \\
                                        & $S_{v}/S_{vi}$ & 179.1          
\end{tblr}
\end{table}

\bibliographystyle{apsrev4-1}
\bibliography{references}